# Quantitative quantum mechanical approach to SABRE hyperpolarization at high magnetic fields

**Stephan Knecht[†,‡] and Konstantin L. Ivanov[§,*]**

[†]*Eduard-Zintl Institute for Inorganic and Physical Chemistry, TU Darmstadt, Alarich-Weiss-Str. 8, D-64287 Darmstadt, Germany*

[‡]*Deptartment of Radiology, Medical Physics, Medical Center – University of Freiburg, Faculty of Medicine, University of Freiburg, Germany*

[§]*International Tomography Center, Siberian Branch of the Russian Academy of Science, Novosibirsk, 630090 (Russia) and Novosibirsk State University, Novosibirsk, 630090 (Russia)*

\* Corresponding author, email: ivanov@tomo.nsc.ru

**Abstract**

A theoretical approach is proposed for quantitative modeling of SABRE (Signal Amplification By Reversible Exchange) experiments performed at a high magnetic field of an NMR spectrometer. SABRE is a method, which exploits the spin order of parahydrogen (the $H_2$ molecule in its nuclear singlet state) for hyper-polarizing the spins of various substrates to enhance their NMR signals. An important feature of SABRE is that the substrate is not modified chemically: instead spin order transfer takes place in a transient complex with parahydrogen. In high-field SABRE experiments, such a transfer is achieved by using suitable NMR excitation schemes. The approach presented here can explicitly treat the spin dynamics in the SABRE complex as well as the kinetics of substrate exchange (between the free and bound form) and complex interplay of spin evolution and chemical processes. One more important effect included in the model is the alteration of the spin state of parahydrogen giving rise to the formation of anti-phase spin order from the initial singlet order. Such a treatment enables a detailed analysis of known high-field SABRE schemes, quantitative comparison with experiments and elucidation of the key factors that limit the resulting NMR signal enhancement.

# I. Introduction

SABRE (Signal Amplification By Reversible Exchange) polarization[1-3] has drawn significant attention during the last decade as a promising and cost-efficient method to enhance weak NMR signals. SABRE exploits parahydrogen ($p$H$_2$, the H$_2$ molecule in the singlet nuclear spin state) as a source of non-thermal spin order. In a SABRE experiment, $p$H$_2$ and a to-be-polarized substrate bind transiently to a Ir-based organometallic complex (SABRE complex), where spin order transfer takes place. After spin order transfer the substrate molecule gets polarized and dissociates from the complex. An advantage of SABRE in contrast to traditional hydrogenative parahydrogen Induced Polarization (PHIP)[4-7] is that the substrate is not modified chemically and is therefore not consumed. As a consequence, hyperpolarization can be generated again by flushing the $p$H$_2$ gas through solution and performing spin order transfer to obtain detectable substrate magnetization.

Typically, in SABRE experiments polarization transfer occurs in a "spontaneous" way, i.e., without radiofrequency (RF) excitation, when the spin system is placed at a low magnetic field, which is set to 5-15 mT to polarize protons[8, 9] and to µT fields (or sub-µT fields) to polarize hetero-nuclei,[10-14] such as $^{13}$C, $^{15}$N, $^{31}$P. Hence, for high-resolution NMR detection of SABRE-polarized molecules it is necessary to transport the sample between the low polarization field and the high NMR detection field. Fast and reproducible sample transport (requiring field cycling) can be technically demanding. Furthermore, it is a limiting factor when it comes to the study of fast reaction kinetics and signal averaging. As of to date, several techniques have been proposed during the last years, which exploit special NMR pulse sequences, rely either on spin mixing at level crossings occurring in the rotating frame[15-20] or on coherence transfer, for spin order transfer at high fields[21-26]. "Spontaneous" polarization transfer in high-field SABRE has also been reported[27, 28]; however, such a transfer, relying on cross-relaxation, is relatively inefficient.

In this contribution we seek to introduce a theoretical treatment of SABRE at high $B_0$ fields and analyze the efficiency of previously designed spin order transfer schemes. An important feature of our treatment is that we explicitly consider both spin dynamics of the SABRE complex and chemical kinetics of substrate exchange between the free and complex-bound form. Such a treatment is analogous to the approach proposed before[29] to spontaneous polarization transfer in SABRE. Special consideration is given to the change of spin order of H$_2$ resulting from singlet-triplet mixing[30], see explanation below. The effect of this spin order change is examined for several transfer schemes in order to assist researchers in optimizing such schemes to their experimental conditions. We also believe that our approach will be useful to understand the source of this mixing in future investigations. By considering the chemical kinetics, spin order conversion and initial spin state of H$_2$ we can make realistic predictions, which should be helpful for application of high-field SABRE techniques.

Pulse sequences have become an important tool at high field analysis; they are routinely employed[24, 31, 32] by Eshuis, Tessari and co-workers for trace analysis at high fields. Such pulse sequences have also proven to be an easy path to high SABRE polarization of X-nuclei as well as a straightforward tool to gain inside into relevant kinetic parameters, such as exchange and decay rates. Here we aim to provide an analysis of INEPT-based NMR pulse sequences[22, 23] designed for high-field SABRE, but other sequences can be treated the same way. In addition, we treat here pulse sequences, which exploit spin mixing at level crossings in the rotating frame that are mentioned above.

We choose to examine two model systems for each class of transfer schemes in order to exemplify our approach. This way of analysis can be extended to a multitude of schemes and problems involving several substrates and complexes. Such theoretical considerations are helpful for development of new polarization transfer techniques (as well as their adaptation to the unique conditions of SABRE systems) and for estimating their actual efficiencies. Our recent development of SLIC-SABRE[18] and repolarization[23] are examples of such a theory-assisted development.

## II. Theory

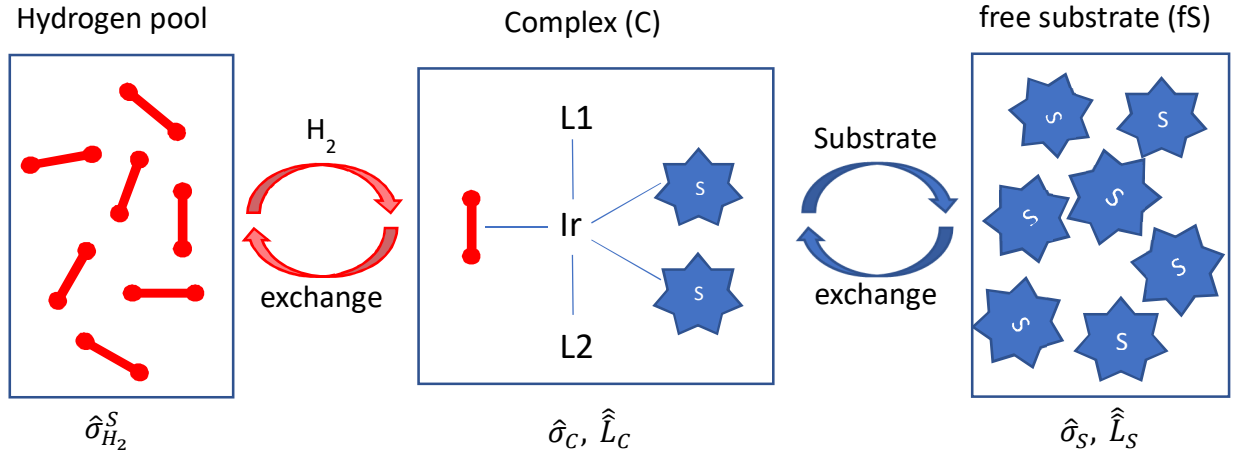

**Scheme 1.** Scheme of the processes giving rise to the formation of SABRE hyperpolarization. The SABRE complex (C) binds dihydrogen and the substrate (S). When the H$_2$ molecule is prepared in a non-equilibrium spin state given by the density matrix $\hat{\sigma}_{H_2}^S$ polarization transfer in the complex takes place giving rise to hyperpolarization of bound S. Exchange between the free and bond forms of S gives rise to polarization transfer to the free substrate (fS) pool. The spin states of C and fS are described by the density matrices $\hat{\sigma}_C$ and $\hat{\sigma}_S$, respectively; their spin evolution is given by the Liouville operators $\hat{\hat{L}}_C(t)$ and $\hat{\hat{L}}_S(t)$, respectively.

In this section we present a closed set of kinetic equations for the density matrix of the free substrate (fS) and the SABRE complex (C), see **Scheme 1**. These equations explicitly take into account the spin dynamics of the substrate in its free and bound form, as well as chemical exchange between these species. Such equations generalize the treatment proposed previously[29] for spontaneous polarization transfer. We also discuss algorithms for solving the proposed set of equations; these algorithms remain valid for time-dependent Hamiltonian of the spin system as well. In practice, the time dependence arises from application of NMR pulses for manipulating spin order transfer. In view of recent high-field PHIP/SABRE works[18, 28, 30], an important issue is the spin order of H$_2$. Although PHIP/SABRE experiments rely on using $p$H$_2$, i.e., the initial state of H$_2$ is the singlet state, $|S\rangle$, in solution containing a PHIP/SABRE complex fast spin mixing occurs, which gives rise to formation of H$_2$ in the central triplet state, $|T_0\rangle$. Hence, the initial state of H$_2$ in solution, which is the source of the non-thermal spin order exploited to polarize the substrate, can be different from the singlet order.

### A. Kinetic equations

As a starting point for our treatment, we write down coupled differential equations for the density matrices of the free substrate and the SABRE complex, $\hat{\sigma}_S$ and $\hat{\sigma}_C$, respectively. These equations are as follows:

$$\frac{d\hat{\sigma}_S}{dt} = \hat{\hat{L}}_S(t)\hat{\sigma}_S - W_a\hat{\sigma}_S + k_d \text{Tr}_{H_2}\{\hat{\sigma}_C\} \tag{1a}$$

$$\frac{d\hat{\sigma}_C}{dt} = \hat{\hat{L}}_C(t)\hat{\sigma}_C - k_d\sigma_c + W_a\{\hat{\sigma}_S \otimes \hat{\sigma}_{H_2}^S(t)\} \tag{1b}$$

Here the terms containing $\hat{\hat{L}}_S(t)$ and $\hat{\hat{L}}_C(t)$ (Liouville operators of fS and the complex) describe the spin dynamics of fS and the complex. They have the form

$$\widehat{\hat{L}}_{S,C}(t) = -i\widehat{\hat{H}}_{S,C}(t) + \widehat{\hat{R}}_{S,C}$$

where $\widehat{\hat{H}}_{S,C}$ and $\widehat{\hat{R}}_{S,C}$ stand for dynamic and stochastic evolution, respectively. The terms containing $W_a$ (association rate of the complex) and $k_d$ (dissociation rate of the complex) describe chemical exchange of the free and bound (bound substrate, bS) forms of the substrate. In equation (1a) association is accounted for as a simple exchange at a rate $W_a$, whereas in the dissociation term we also remove extra-dimensionality of the density matrix $\hat{\sigma}_C$ by tracing out spins, which are present in the complex being absent in fS. In the simplest case considered here, we assume that these additional spins are only the protons of H$_2$. In the second equation, dissociation is described as decay at a rate $k_d$, while the association term also increases the dimensionality of the density matrix, which is done by taking the direct product of $\hat{\sigma}_S$ and the density matrix of H$_2$ in solution, here denoted as $\hat{\sigma}_{H_2}^S$. Detailed explanation for such a structure of equations has been given in our previous paper[29].

The set of equations introduced here should be solved in the Liouville space, where each density matrix should be presented as a column-vector. In addition, it is convenient to introduce a generalized density matrix of the entire spin system as a column-vector $\hat{\sigma}$ containing both $\hat{\sigma}_S$ and $\hat{\sigma}_C$. Equation 1 can then be re-written as a single matrix-form equation:

$$\frac{d}{dt}\begin{pmatrix}\hat{\sigma}_S(t)\\ \hat{\sigma}_C(t)\end{pmatrix} = \frac{d\hat{\sigma}(t)}{dt} = \hat{\hat{A}}(t)\begin{pmatrix}\hat{\sigma}_S(t)\\ \hat{\sigma}_C(t)\end{pmatrix} = \hat{\hat{A}}(t)\hat{\sigma}(t) \qquad (2)$$

with the generalized superoperator

$$\hat{\hat{A}}(t) = \begin{pmatrix}\widehat{\hat{L}}_S(t) - W_a\hat{\hat{1}}_S & k_d\hat{\hat{S}}^{Tr}\\ W_a\hat{\hat{S}}^{\otimes}(t) & \widehat{\hat{L}}_C(t) - k_d\hat{\hat{1}}_C\end{pmatrix} \qquad (3)$$

Here $\hat{\hat{1}}_S$ and $\hat{\hat{1}}_C$ are the unity super-operators in the corresponding sub-spaces of spin states. For the off-diagonal elements in $\hat{\hat{A}}(t)$, which describe transitions between fS and bS, we use the following short-hand notations:

$$\hat{\hat{S}}^{Tr}\hat{\sigma}_C = \mathrm{Tr}_{H_2}\{\hat{\sigma}_C\}, \qquad \hat{\hat{S}}^{\otimes}(t)\hat{\sigma}_S = \hat{\sigma}_S \otimes \hat{\sigma}_{H_2}^S(t)$$

In contrast to our previous work[29] here we consider explicitly the time dependence of $\hat{\hat{A}}(t)$, which originates from (i) the time dependence of the Hamiltonians $\widehat{\hat{H}}_{S,C}$ and (ii) the time dependence of the H$_2$ spin state. Hence, the scheme for numerical solution of the generalized equation for $\hat{\sigma}(t)$ needs to be updated.

In the presence of RF fields used for NMR excitation the spin Hamiltonian becomes time-dependent due to the time dependence of the external RF-field. Let us consider the situation where we have two species of spins in the system ($I$ and $S$, e.g., protons and $^{15}$N spins) and RF excitation is performed on the $I$-spins. Such a consideration is sufficiently general to cover all existing RF-excitation schemes used in SABRE. Hereafter, we consider $N$ $I$-spins and $M$ $S$-spins; the $I$-spins are always protons with spins 1 and 2 being the protons originating from $p$H$_2$. If the RF-field is circularly polarized, the Hamiltonian in the lab frame can be written as follows:

$$\hat{H} = -\sum_{i=1}^{N}\omega_i^I\hat{I}_{iz} + \sum_{i<j}2\pi J_{ij}(\hat{\mathbf{I}}_i \cdot \hat{\mathbf{I}}_j) - \sum_{i=1}^{N}\omega_{1I}\{\cos(\omega_{rf}^I t)\hat{I}_{ix} + \sin(\omega_{rf}^I t)\hat{I}_{iy}\}$$
$$-\sum_{k=1}^{M}\omega_k^S\hat{S}_{kz} + \sum_{k<m}2\pi J_{km}(\hat{\mathbf{S}}_k \cdot \hat{\mathbf{S}}_m) - \sum_{k=1}^{M}\omega_{1S}\{\cos(\omega_{rf}^S t)\hat{S}_{kx} + \sin(\omega_{rf}^S t)\hat{S}_{ky}\} \qquad (4)$$
$$+\sum_i\sum_k 2\pi J_{ik}(\hat{\mathbf{S}}_k \cdot \hat{\mathbf{I}}_i)$$

Here we introduce the NMR frequencies of spins as $\omega_i^I = \gamma_I B_0(1 + \delta_i^I)$ and $\omega_k^S = \gamma_S B_0(1 + \delta_k^S)$ (with $\gamma_{I,S}$ being the corresponding gyromagnetic ratios, $\delta_i^I$ and $\delta_k^S$ being the chemical shifts), $J_{ij}$ are the scalar coupling constants measured in Hz, $\omega_{1I} = \gamma_I B_{1I}$ and $\omega_{1S} = \gamma_S B_{1S}$ stand for the strength of the oscillating RF-fields $B_{1I}$ and $B_{1S}$, measured in the frequency units. Generally, the NMR parameters are different for fS and bS; in the Hamiltonian of the SABRE complex it is necessary to consider additional nuclei (at least, the spins of H₂). Commonly, solution of the problem is simplified by transforming the Hamiltonian in the rotating frame of reference (the frequency of the frame is given by $\omega_{rf}^I$ for $I$-spins and by $\omega_{rf}^S$ for $S$-spins). After transformation, we arrive at the following Hamiltonian:

$$\widetilde{\hat{H}} = -\sum_{i=1}^{N} \Omega_i^I \hat{I}_{iz} + \sum_{i<j} 2\pi J_{ij}(\hat{\mathbf{I}}_i \cdot \hat{\mathbf{I}}_j) - \sum_{i=1}^{N} \omega_{1I} \hat{I}_{ix}$$
$$-\sum_{k=1}^{M} \Omega_k^S \hat{S}_{kz} + \sum_{k<m} 2\pi J_{ij}(\hat{\mathbf{S}}_k \cdot \hat{\mathbf{S}}_m) - \sum_{k=1}^{M} \omega_{1S} \hat{S}_{kx} + \sum_i \sum_k 2\pi J_{ik} \hat{S}_{kz} \hat{I}_{iz} \qquad (5)$$

In the rotating frame, the frequencies of the $I$-spins and $S$-spins become $\Omega_i^I = \omega_i^I - \omega_{rf}^I$ and $\Omega_k^S = \omega_k^S - \omega_{rf}^S$, while the $B_1$-terms become time-independent. As usual, in the terms, which stand for couplings between $I$-spins and $S$-spins we omit all rapidly oscillatory contributions (associated with $\hat{I}_{ix}$ and $\hat{I}_{iy}$) and write the coupling terms as $2\pi J_{ik} \hat{S}_{kz} \hat{I}_{iz}$. At the same time, coupling terms within the groups of $I$-spins and $S$-spins remain the same as in the lab frame. When the RF-field is applied on only one of the two channels, frame rotation should be performed only for the spins of the corresponding type.

Solution of the problem depends on the RF-excitation scheme employed. When a continuous-wave (CW) RF-field is applied, consideration of the spin dynamics in the rotating frame is equivalent to that done earlier in the absence of RF-fields[29]. When the RF-field strength (or phase) is time-dependent, we need to integrate eq. (2) explicitly. The time-dependence of the spin state of H₂ in solution, i.e., the time dependence of $\hat{\sigma}_{H_2}^S$, also results in a time-dependent propagator $\hat{\hat{A}}(t)$ in eq. (3). We choose to solve the resulting equations in a stepwise manner, each step chosen on a timescale short enough, so that $\hat{\hat{A}}(t) \cong \hat{\hat{A}}(t + dt)$. In the case of short resonant pulses, the effect of all pulses was considered as instantaneous rotations of the corresponding spins by the pulse tip angle. Between the pulses the Hamiltonian is time-independent, which makes the calculation simpler.

## B. Spin order of H₂ in the SABRE system

Spin order of H₂ in solution is determined by two factors: external supply of hyperpolarized H₂ (as described by the density matrix $\hat{\sigma}_{H_2}^{ex}$) and evolution of the spin order of H₂ in solution. Hence, we introduce the density matrix $\hat{\sigma}_{H_2}^S$, which describes the state of H₂ in solution. To describe the spin evolution of H₂, one should note that H₂ undergoes constant exchange between the free and complex-bound forms; when H₂ binds to the SABRE complex spin conversion becomes operative, as explained below. We can describe the evolution of the spin state of H₂ in solution by the following phenomenological equations:

$$\frac{d}{dt} \hat{\sigma}_{H_2}^S = \left\{\hat{\hat{L}}_{H_2} - \hat{\hat{K}}_{ex}\right\} \hat{\sigma}_{H_2}^S + \hat{\hat{K}}_{ex} \sigma_{H_2}^{ex} \qquad (6)$$

Where $\hat{\hat{L}}_{H_2}$ is a generalized Liouvillian describing the evolution of solution state of hydrogen (including the time evolution of the spin state caused by the exchange with the SABRE complex) and $\hat{\hat{K}}_{ex}$ is the superoperator, describing exchange with the externally supplied hydrogen.

To describe the actual spin state of H₂ in solution we take into account $S$-$T_0$ mixing, which is known to occur when H₂ binds to the SABRE complex[30, 33, 34]. The physical origin of such mixing is as follows. When H₂ binds to the complex, the two protons become non-equivalent (in some cases, chemically non-equivalent, in some cases, only weakly, i.e., magnetically non-equivalent). Hence, singlet-triplet transitions in H₂ become allowed. At high magnetic fields, these transitions occur predominantly between the $|S\rangle$ and $|T_0\rangle$ states, since the $|T_\pm\rangle$ states are way too far in energy. The $S$-$T_0$ transitions give rise to alteration of the spin order of H₂ in solution; specifically, H₂ in the triplet state (orthohydrogen, oH₂) is formed with the $|T_0\rangle$ state enriched as compared to the $|T_\pm\rangle$ states. Such a spin order of H₂ has very specific spectral manifestation[28, 30], therefore, it can be unequivocally identified by NMR spectroscopy.

Hence, for developing the theoretical treatment of high-field SABRE we need to take into account $S$-$T_0$ mixing. To do so, we introduce the density matrices of a spin pair in the $|S\rangle$ state and in the $|T_0\rangle$ state, as well as their superposition:

$$\hat{\rho}_S = \frac{1}{4}\hat{E} - \hat{I}_{1x}\hat{I}_{2x} - \hat{I}_{1y}\hat{I}_{2y} - \hat{I}_{1z}\hat{I}_{2z}$$
$$\hat{\rho}_{T_0} = \frac{1}{4}\hat{E} + \hat{I}_{1x}\hat{I}_{2x} + \hat{I}_{1y}\hat{I}_{2y} - \hat{I}_{1z}\hat{I}_{2z} \quad (7)$$
$$\hat{\rho}_{S+T_0} = \frac{1}{2}(\hat{\rho}_S + \hat{\rho}_{T_0}) = \frac{1}{4}\hat{E} - \hat{I}_{1z}\hat{I}_{2z}$$

As we pointed out in an earlier work[18, 28], there is evidence, that the spin state (being initially the singlet state) of H₂ is not preserved in the SABRE system, but that there is very rapid $S$-$T_0$ mixing and relaxation. We can account for this behavior by introducing appropriate exchange and (cross) relaxation rates for the relevant populations $P_S$ and $P_{T_0}$. However, it is more convenient to examine the deviations from the thermal equilibrium $\delta P_S = P_S - P_S^{eq}$ and $\delta P_{T_0} = P_{T_0} - P_{T_0}^{eq}$ instead of $P_S$ and $P_T$. The kinetic equations for these quantities are:

$$\frac{d}{dt}\begin{pmatrix}\delta P_S \\ \delta P_{T_0}\end{pmatrix} = \begin{pmatrix}-(R_S + \sigma + K_{ex}) & \sigma \\ \sigma & -(R_{T_0} + \sigma + K_{ex})\end{pmatrix}\begin{pmatrix}\delta P_S \\ \delta P_{T_0}\end{pmatrix} + \begin{pmatrix}K_{ex}\,\delta P_S^{ex} \\ K_{ex}\,\delta P_{T_0}^{ex}\end{pmatrix} \quad (8)$$

Here $R_S$ and $R_{T_0}$ are the relaxation rates of the singlet and $T_0$ triplet state of H₂ in solution, respectively. The rate of mixing between the singlet and $T_0$ state is equal to $\sigma$. The rate $K_{ex}$ denotes the exchange rate with the externally supplied pH₂. The deviations of the populations of the externally supplied H₂ are given by $\delta P_S^{ex}$ and $\delta P_{T_0}^{ex}$, respectively. After bubbling has stopped, the system can be described by eq. (8) without external exchange ($K_{ex} = 0$). Here we decided not to make any assumptions about the relaxation and cross-relaxation parameters in eq. (8), as they will most likely depend strongly on the chemistry of the system under consideration. Instead we choose to explore the following two extreme cases.

The first case we consider is that of no singlet-triplet mixing and fast hydrogen exchange (i.e. $\sigma = 0$ and $k_d, R_S \ll K_{ex}$). This is exactly the assumption used in all previous models including our previous publications. We assume that H₂ in solution is initially formed in the pure singlet state and that the deviation $\delta P_S$ of the population of this state from its equilibrium value decays with a time constant $R_S$. We assume that this decay is non-selective when it comes to the triplet states, meaning that the $|T_+\rangle, |T_-\rangle$ and $|T_0\rangle$ states are always equally populated. It is easy to show that in such a case the density matrix of H₂ in solution is given by:

$$\hat{\sigma}_{H_2}(t) = f_S(t)\,\hat{\rho}_S + \frac{1 - f_S(t)}{3}(\hat{E} - \hat{\rho}_S) \quad (9)$$

Where, the now time-dependent, fraction of singlet state, $f_S$, is introduced. In the present case it is equal to

$$f_S(t) = \delta P_S(t) + \frac{1}{4} = \delta P_S(t = 0) \cdot e^{-t \cdot R_S} + \frac{1}{4}$$

The second limiting case we explore, is that of rapid equilibration of the populations $P_S$ and $P_{T_0}$ (meaning that $\sigma \gg R_s, R_{T_0}$). In such a situation the , $P_S = P_{T_0}$; hence, it is convenient to describe the dynamics of the populations $P_{S+T_0} = \delta P_{S+T_0} + \frac{1}{4}$ of the state $\hat{\rho}_{S+T_0}$ introduced in eq. (7). The resulting density matrix of H₂ is thus as follows:

$$\hat{\sigma}_{H_2}(t) = 2f_{S+T_0}(t)\hat{\rho}_{S+T_0} + \left(\frac{1}{2} - f_{S+T_0}(t)\right)\left(\hat{E} - 2\hat{\rho}_{S+T_0}\right) \quad (10)$$

With

$$f_{S+T_0}(t) = \delta P_{S+T_0}(t=0) \cdot e^{-t \cdot R_{S+T_0}} + \frac{1}{4}$$

## III. Methods
### A. Parameters considered and evaluation of results

Let us discuss the technical and practical aspects of setting up and evaluating the above described equations as well as the parameters used in these simulations. To calculate the evolution of our system in the Liouville space, the Hilbert space density matrix is transformed into a state vector using the following transformation rule, which simply concatenates each row of the density matrix:

$$\hat{\sigma} = \begin{pmatrix} \sigma_{1,1} & \cdots & \sigma_{1,j} \\ \vdots & \ddots & \vdots \\ \sigma_{i,1} & \cdots & \sigma_{i,j} \end{pmatrix} \rightarrow \boldsymbol{\sigma} = \begin{pmatrix} \sigma_{1,1} \\ \vdots \\ \sigma_{1,j} \\ \vdots \\ \sigma_{i,1} \\ \vdots \\ \sigma_{i,j} \end{pmatrix}$$

In order to calculate the time evolution of $\hat{\sigma}$ under the generalized superoperator $\hat{\hat{A}}(t)$ we also need to consider a generalized state vector for both the free substrate (fS) and the SABRE complex (C).

$$\sigma = \begin{pmatrix} \sigma_C \\ \sigma_S \end{pmatrix}$$

In the present work, the spin systems of C and fS are assumed to be non-polarized at the beginning of each experiment, such a treatment is justified by the fact that thermal polarization induced by the static external magnetic field is (in all instances described here) much lower than the hyperpolarization generated by SABRE, having no effect on the spin dynamics. Consequently, the density matrix of our system at time $t = 0$ reads as follows:

$$\hat{\sigma} = \begin{pmatrix} \dfrac{\hat{1}_S}{Dim_S} \\ \dfrac{\hat{1}_C}{Dim_C} \end{pmatrix}$$

If the propagator $\hat{\hat{A}}$ is time-independent the evolution of the density matrix can be easily calculated by integrating eq. (2):

$$\hat{\sigma}(t) = e^{\hat{\hat{A}} \cdot t} \hat{\sigma}(t=0)$$

In cases where $\hat{\hat{A}}(t)$ is time-dependent the propagation is conducted stepwise, choosing each step small enough, such that $\hat{\hat{A}}$ can be considered time independent during each step:

$$\hat{\sigma}(t+t_0) = \left(\prod_{i=1}^{n} \hat{A}(t_0 + i \cdot \Delta t) \cdots \hat{A}(t_0)\right) \hat{\sigma}(t_0), \quad \Delta t = \frac{t}{n}$$

After calculation of the $\hat{\sigma}(t)$ matrix the relevant observable $\hat{O}$ is examined by taking the scalar product in the Liouville space (corresponding to the trace in the Hilbert space):

$$P(\hat{O}) = \langle \hat{O}|\sigma\rangle = trace(\hat{O} \cdot \hat{\sigma})$$

To calculate the coherent evolution under the Hamiltonian of eq. (5) we used the scalar coupling constants and NMR frequencies as listed in **Table 1**. The amplitude ($\omega_1$) and frequency ($\omega_{rf}$) used in the RF-SABRE transfer schemes are reported in the figure captions.

**Table 1**. Parameters of the SABRE complex used in calculations. See text for explanation.

| $J_{ij}$ [Hz] | $H_1$ | $H_2$ | $^{15}N_{e1}$ | $^{15}N_{e2}$ | $H_S$ |
|---|---|---|---|---|---|
| $H_1$ |  | −7 | 0 | −20 | 0.3 |
| $H_2$ |  |  | −20 | 0 | 0 |
| $^{15}N_{e1}$ |  |  |  | −0.4 | – |
| $^{15}N_{e2}$ |  |  |  |  | – |
| $H_S$ |  |  |  |  |  |
|  |  |  |  |  |  |
| $\delta$ [ppm] Complex | −22.7 | −22.7 | 255 | 255 | 8.3 |
| $\delta$ [ppm] Free form | 4.5 | 4.5 | 300 | 300 | 8 |

For the case of RF-SABRE on protons we restrict ourselves to the case of a three-spin system with only one substrate spin. When simulating transfer to hetero-nuclei, e.g., to $^{15}N$ nuclei, we consider a four-spin system comprising the two equatorial $^{15}N$ nuclei, denoted as $^{15}N_{e1}$ and $^{15}N_{e2}$, and two protons of bound $H_2$, denoted as Ir-$H_2$ spins (we neglect coupling to other protons of the substrate). We assume, for simplicity, that both $^{15}N$ nuclei leave the complex simultaneously. When modeling polarization transfer to protons, we assume that there are no $^{15}N$ nuclei (because of their low natural abundance) and consider a three-spin system of the two Ir-$H_2$ protons and a single substrate proton, denoted as $H_S$. The Hamiltonian in the Hilbert space is readily implemented numerically by replacing the constants of eq. (4) with their numerical values and the spin operators with their matrix representations, respectively. In the Hilbert space, the time dependence of the density matrix under coherent evolution is given by the well-known Liouville-von Neumann equation:

$$\frac{d}{dt}\hat{\sigma} = -i[\hat{H}, \hat{\sigma}]$$

In the Liouville space, this equation is written by defining the appropriate super-operator, for the commutator term:

$$\frac{d}{dt}|\hat{\sigma}\rangle = -i\hat{\hat{H}}_C|\hat{\sigma}\rangle$$

Where $\hat{\hat{H}}_C$ is the so-called right-hand side commutation super-operator of $\hat{H}$:

$$\hat{\hat{H}}_C = \hat{H} \otimes \hat{1} - \hat{1} \otimes \hat{H}$$

For calculation of the relaxation super-operator we used the same model of random fluctuating fields in the extreme narrowing limit[29, 35, 36] as we did in earlier works. This model is sufficient for our purposes and rather simple to define: the only parameters needed for calculation of the relaxation super-operator are the high-field relaxation rates $R_1 = \frac{1}{T_1} = \frac{1}{T_2} = R_2$ of each spin in the system. All the above parameters are reported in the figure captions. In cases where we consider the time dependence of the H$_2$ spin order in solution explicitly, compare eqs. (8–10), the relaxation rate $T_1^{H_2}$ of this spin order is given as well. One may choose to use a treatment, explicitly considering relaxation inducing interactions (e.g. dipolar interaction) or alternatively a purely empirical relaxation treatment may be employed.

Because chemical exchange is in a steady-state (i.e., the concentrations of fS and C do not change) there is a straightforward relationship between the exchange rates and the concentrations of the complex and fS, namely, $\frac{[C]}{[fS]} = \frac{k_d}{W_a}$. This means that the parameters governing the chemical exchange in our model are determined by specifying the dissociation rate $k_d$ and the concentrations $[C] = [bS]$ and $[fS]$. The linear operators for the partial trace and direct product used here are the same as in our previous work.

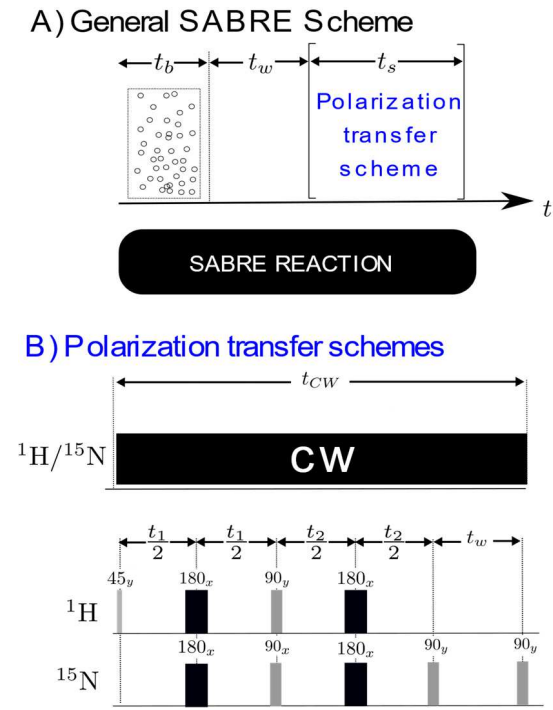

**Figure 1**. A) General schematic representation of the SABRE reaction and polarization transfer process at high magnetic fields considered in the simulation. Firstly, $p$H$_2$ is bubbled through the solution directly at a high spectrometer field. This is usually followed by a delay needed to remove bubbles from the solution. Secondly, a suitable polarization transfer scheme is applied using NMR excitation. B) Schematic depiction of the two transfer schemes selected for analysis in this work. (Top): simplified version of a LIGHT-SABRE or RF-SABRE experiment: polarization transfer is performed in the presence of a CW RF-field. (Bottom): SABRE-INEPT sequence.

## B. Simulation Schemes

In this work, we consider the following scheme of a SABRE experiment (**Figure 1A**). We assume that spin order of H$_2$ is prepared by $p$H$_2$ bubbling during a time $t_b$; here we do not consider the spin dynamics of $p$H$_2$ during bubbling and only assume that it generates the initial condition for $\hat{\sigma}_{H_2}^S$. After that, polarization transfer is taking place in the SABRE complex during a time period $t_s$; simultaneously polarization is transferred to fS via chemical exchange. After applying the transfer scheme, a waiting time $t_w$ is introduced in some experiments to enable polarization transfer from bS to fS via chemical exchange. Here we simulate all three aspects of the SABRE reaction scheme: polarization transfer in the SABRE complex, exchange of $H_2$, which alters the spin order given by $\hat{\sigma}_{H_2}^S$, and substrate exchange between bS and fS. In simulations, we consider explicitly both the spin dynamics and chemical exchange during the polarization transfer step and during the period of free evolution. The exchange process between bS and fS is explicitly taken into account at all times. As far as the $H_2$ exchange during the polarization transfer is concerned, in some cases it is considered explicitly, while in most cases we modify the density matrix $\hat{\sigma}_{H_2}^S$ and keep it time-independent.

As far as the polarization transfer schemes are concerned, we consider two distinctly different cases, see **Figure 1B)**.

In the first case, a single CW pulse is applied for polarization transfer. The transfer mechanism in this case is based on spin order transfer at level crossings in the rotating frame. After writing down the Hamiltonian in the rotating frame, we obtain time-independent Liuoville operators describing the spin evolution and hence the time-independent superoperator $\hat{\hat{A}}$. When the time dependence of $\hat{\sigma}_{H_2}^S$ is also omitted, we can use the same scheme for numerical solution as previously[29]. A method, which makes use of long CW-pulses, is named RF-SABRE[15, 16]. In this technique a single NMR pulse is applied on the proton channel with the parameters (i.e., amplitude and frequency) set such that level crossing conditions are fulfilled. We also apply the same calculation scheme to model polarization in LIGHT-SABRE[17] experiments. In this case, polarization transfer is driven by a long CW-pulse on the $^{15}N$ channel; the parameters of the pulse are set such that the level-crossing conditions are fulfilled in the rotating frame. To be precise, we mention that this is a simplified description of the LIGHT-SABRE experiment: in real experiment the CW-pulse is followed by a $\frac{\pi}{2}$ pulse on the nitrogen channel required to convert transversal polarization into longitudinal polarization. Additionally, RF-pulses are selective and excite only bS. Here, for simplicity, we assume that the NMR frequencies of bS and fS are identical. However, for our purposes this small modification is not essential.

In the second case, we consider multi-pulse sequences, e.g., an INEPT-type pulse sequence, which utilizes coherence transfer (here we consider the SABRE-INEPT pulse scheme used previously[21-23]). Effects of all pulses are considered as instantaneous rotation of spins by the corresponding angles. Between the pulses the spins freely evolve and chemical exchange is going on. It should be noted, that all the effects described in the theory section are present in all schemes considered here. For example, the difference in the NMR frequencies of bS and fS will lead to dephasing of the transversal hyperpolarization upon exchange and the waiting time after bubbling, which (usually introduced for the simple technical reason of removing bubbles from the solution to avoid line broadening by susceptibility shifts) will lead to reduced polarization when INEPT-type sequences are used. However, for the sake of clarity, we choose not to present an analysis of all these aspects here (even though they are implemented in our models) but prefer to focus on key features of each method.

## IV. Results and discussion
### A. Spin order of $H_2$

Before modeling SABRE-derived NMR enhancements, let us consider the spin order of the primary source of polarized spins here being $H_2$ in solution. **Figure 2** shows the dependence of the singlet-order and $T_0$ order of $H_2$ depending on the rate of external supply of $pH_2$. The results presented here come from the steady-state solution of eq. 8. One can clearly see that only when $K_{ex}$ is much larger than the rates of relaxation and $S$-$T_0$ conversion, the singlet-state population, $P(S)$, is considerably larger than the $T_0$-state population, $P(T_0)$. In the opposite case of small $K_{ex}$ values we obtain $P(S) \approx P(T_0)$ meaning that the spin order of H$_2$ is no longer the singlet order but rather the pure anti-phase spin order: $\hat{\sigma}^S_{H_2} = \frac{1}{4}\hat{E} - \hat{I}_{1z}\hat{I}_{2z}$.

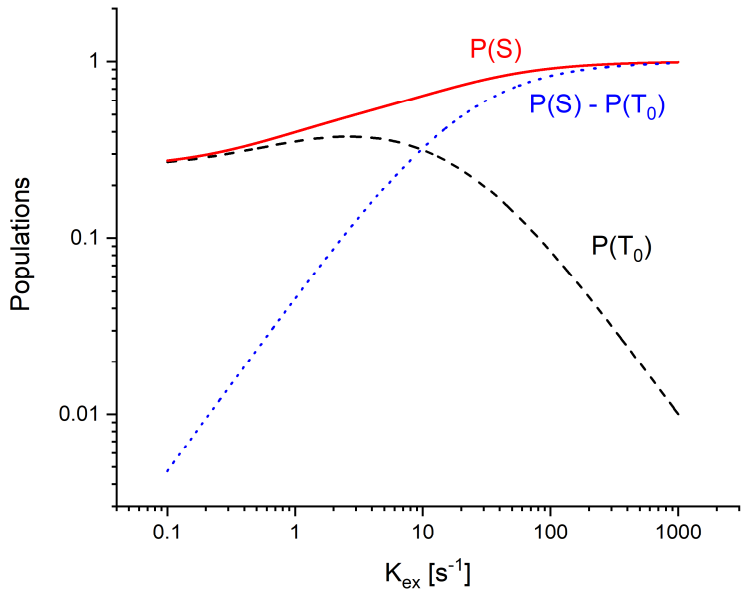

**Figure 2.** Spin order of H$_2$ as a function of $K_{ex}$, the rate of external $p$H$_2$ supply. It is noteworthy, that there are two regimes: when mixing and relaxation are fast compared to the efficiency of external $p$H$_2$ supply there is almost no overpopulation of the singlet state as compared to $|T_0\rangle$ and the resulting spin order is $\hat{I}_{1z}\hat{I}_{2z}$, see eq. (2). When $p$H$_2$ is supplied significantly faster than the mixing and relaxation rates, the resulting state of $p$H$_2$ is the singlet spin order. Parameters used for simulations are $R_S = 1$ s$^{-1}$, $R_{T_0} = 1$ s$^{-1}$ and $\sigma = 10$ s$^{-1}$.

Hence, we can derive two important intermediate conclusions: (i) the initial conditions for transfer schemes are altered; (ii) relaxation of hydrogen spin order in solution is increased due to the increased singlet state leakage into the $|T_0\rangle$ state. Both factors need to be taken into account when designing[18] high-field SABRE transfer schemes and when modelling SABRE-derived polarization at high fields.

## B. Polarization transfer in RF-SABRE

RF-SABRE[15] represents the first successful example of proton polarization at high magnetic fields. Despite successful implementation of the method, one should note that the performance of RF-SABRE was considerably worse than that of the standard SABRE experiments with polarization transfer performed at low magnetic fields. Here we aim to explain the reason for the difference in the performance of high-field and low-field SABRE experiments. As we demonstrate below, the performance of RF-SABRE is dependent on the initial spin state of H$_2$, being approximately an order of magnitude better for singlet-state preparation of H$_2$.

The results of modeling the RF-SABRE efficiency are presented in **Figure 3**. One can see that the matching conditions do not depend on the spin-order effects: the dependence of polarization on the RF-field parameters, i.e., on $\omega_1$ and $\omega_{rf}$ is similar in both cases. However, the efficiency is decreased by more than a factor of 3 for the anti-phase order as compared to the pure singlet order of H$_2$.

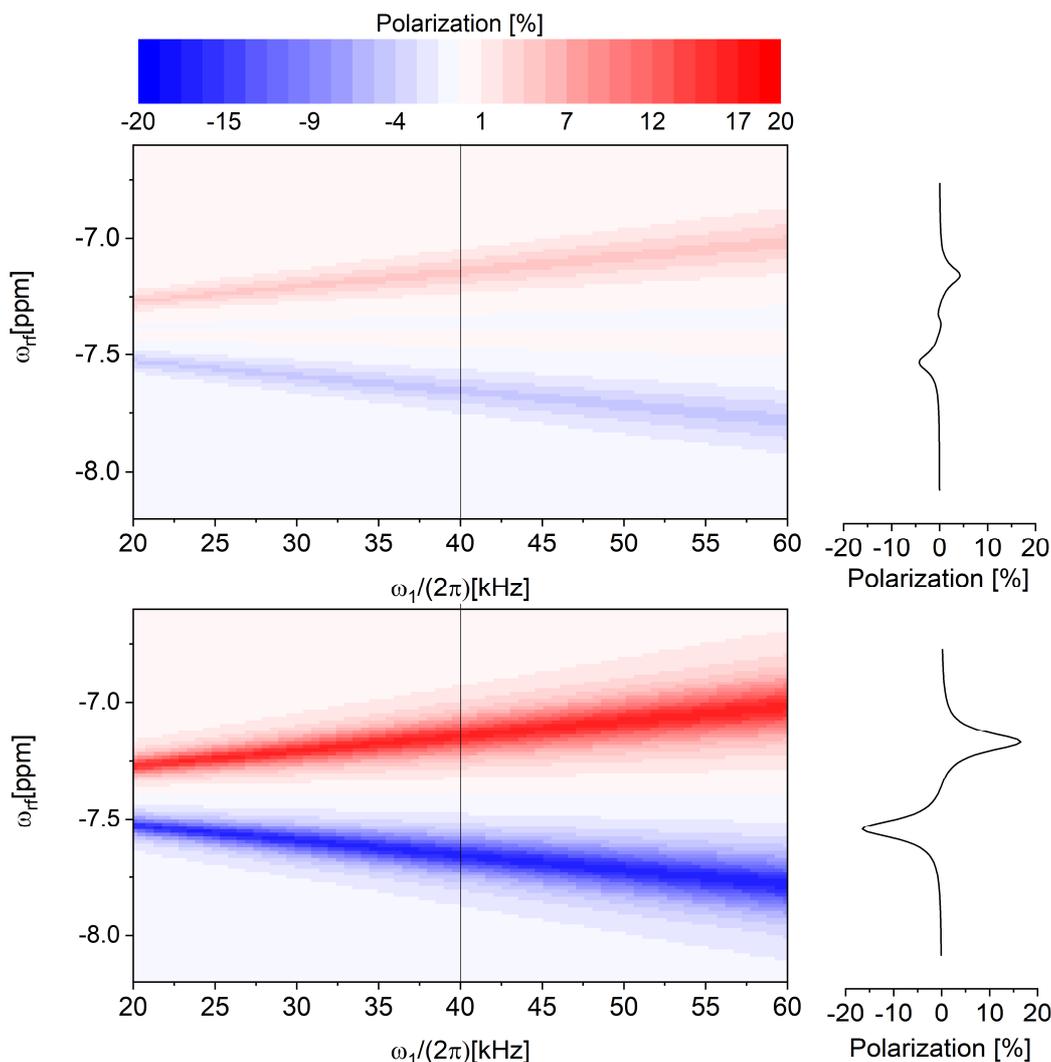

**Figure 3**. Simulations of the free-substrate polarization dependence in the RF-SABRE case; here the dependence of the resulting polarization on the RF frequency $\omega_{rf}$ and amplitude $v_1$ is shown. Polarization is shown in percent. Simulation parameters: $k_d = 10, \frac{[fS]}{[bS]} = 30, t_b = 10$ s, $k_d = 10, \frac{[fS]}{[bS]} = 30, t_b = 10$ s, $T_1^{fS} = 30$ s, $T_1^{H_2} = 1$ s, $T_1^{bS} = 3$ s. Top: spin order of H$_2$ is the anti-phase order. Bottom: spin order of H$_2$ is the singlet order.

It becomes apparent from **Figure 3** that the resonance frequency of the RF-field needs to be precisely set to ensure proper performance of the RF-SABRE method and therefore, hydrogen bubbles need to be removed from the sample prior to the application of the RF-field. Thus, the supply of $p$H$_2$ in the solution is limited. Since the H$_2$ spin order decays back to thermal equilibrium either by participating in the SABRE process, or by spin-lattice relaxation, or by a combination of both. In **Figure 4** this situation is explored by introducing a lifetime of the spin-order in solution, compare eqs. (9) and (10). Here an infinite lifetime means that for the entire duration of the polarization process the free hydrogen pool is comprised entirely of molecules in the singlet state or of molecules having anti-phase two-spin order. With higher decay rates of the spin-order in solution, the efficiency and optimal duration of the CW-field decrease. We suggest two possibilities to address this issue: firstly, repolarization cycles may be introduced, as we did in our previous work with the SABRE-INEPT method[23]. Secondly, alternative routes of supplying $p$H$_2$ to the solution, as have been explored e.g. by Münneman and coworkers[37], offer a possibility to supply fresh $p$H$_2$ without disturbing the NMR lines. It is worth mentioning, that at this point our model does not include any contributions of polarized H$_2$, which may interfere negatively with the SABRE process.

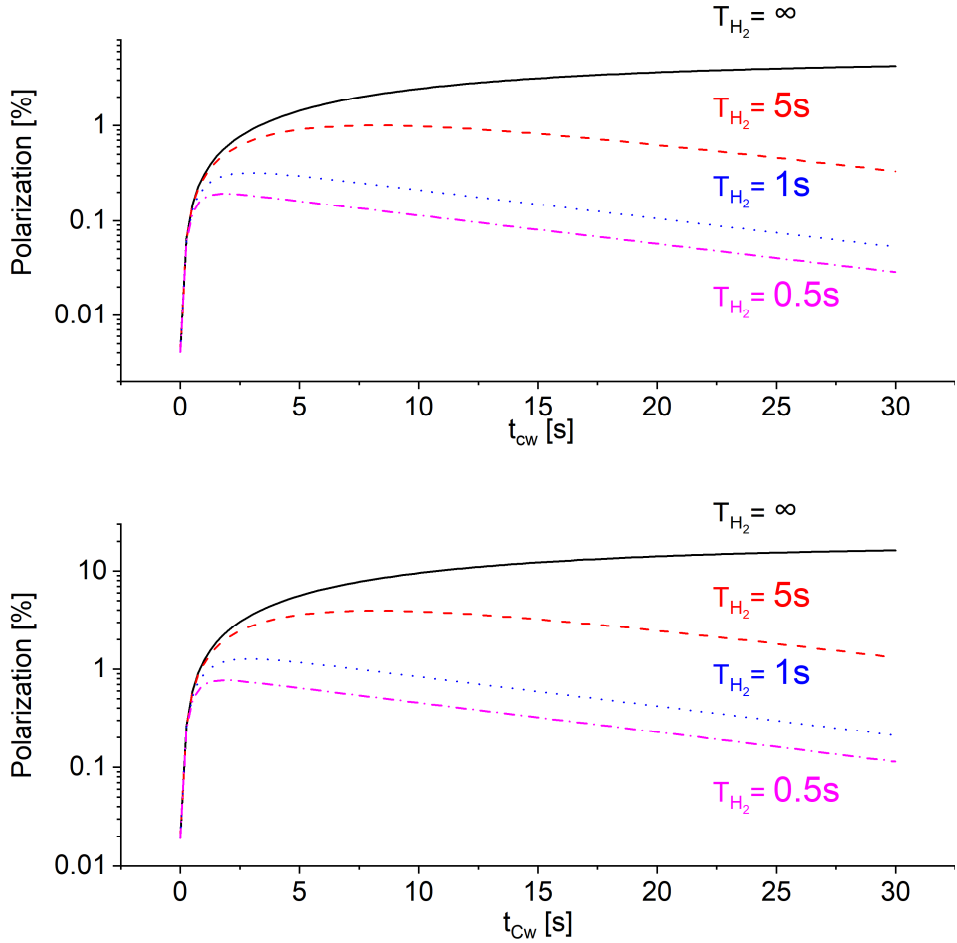

**Figure 4**. Simulations of the free-substrate polarization dependence of RF-SABRE on the time of RF irradiation ($t_{cw}$) for different lifetimes of the hydrogen spin order, $T_s = 1/R_s$. Polarization is shown in percent. Parameters: $k_d = 10$, $\frac{[fS]}{[bS]} = 30$, $t_b = 10$ s, $T_1^{fS} = 30$ s, $T_1^{IrH_2} = 1$ s, $T_1^{bS} = 1$ s. Top: results for the S-T$_0$ limit. Bottom: results for the singlet limit.

## C. LIGHT-SABRE

Using our simulation method, we also consider LIGHT-SABRE and analyze its performance depending on the spin state of H$_2$. As follows from our analysis, the performance of the method strongly depends on the initial spin state of H$_2$. The reason is that LIGHT-SABRE exploits the population difference between the $|S\rangle$ and $|T_0\rangle$ states of H$_2$, which is large for the singlet order and diminishes for anti-phase spin order. Hence, in LIGHT-SABRE almost no polarization is expected when singlet-triplet mixing is operative. This is in accordance with our previous experimental findings[18].The LIGHT-SABRE polarization dependence on the RF-field parameters has been simulated and the results are reported in **Figure 5**.

Obviously when bS and fS have different resonance frequencies (which is the usual case rather than an exception) any polarization generated in the bound form, which acquires a phase difference, would than dephase because of the random nature of the exchange process. Thus an experimentalist is offered several ways to remedy this problem: generate z-magnetization by pulsing, RF-ramping or choosing a resonance condition which directly generates z-magnetization. These techniques have been studied extensively in previous works[18, 38]. It also becomes clear, that a continuously applied CW-field is beneficial in terms of efficiency compared to a scheme were transfer is conducted in a pulsed fashion. The effect of spin-order alteration causing by $S$-$T_0$ mixing has been explored in detail in our recent work[18], which used the formalism presented here for simulations.

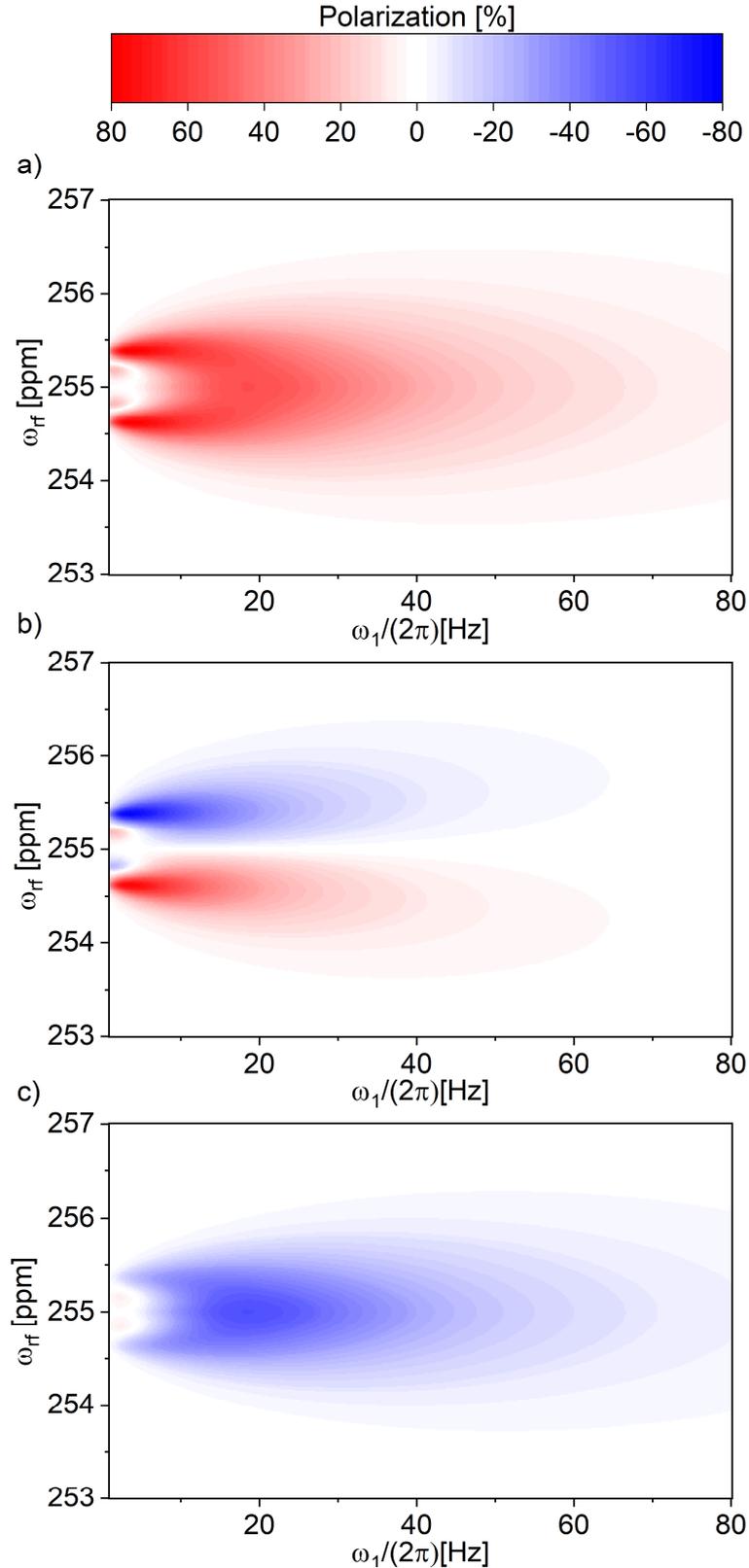

**Figure 5**. Simulations of the fS polarization in LIGHT-SABRE; here the dependence of polarization on the RF frequency $\omega_{rf}$ and amplitude $\omega_1$ is shown. Parameters: $k_d = 10$, $\frac{[fS]}{[bS]} = 30$, $t_b = 60$ s, $T_1^{IrH_2} = 1$ s, $T_1^{fS} = 30$ s and $T_1^{bS} = 3$ s. Relaxation of hydrogen in solution is not considered. Figure shows the singlet-state limit, note that negligible polarization is generated for the anti-phase spin order of H$_2$. A) fS Polarization along the effective field, in the case were fS and bS have the same chemical shift. B) longitudinal fS polarization C) transversal fS polarization.

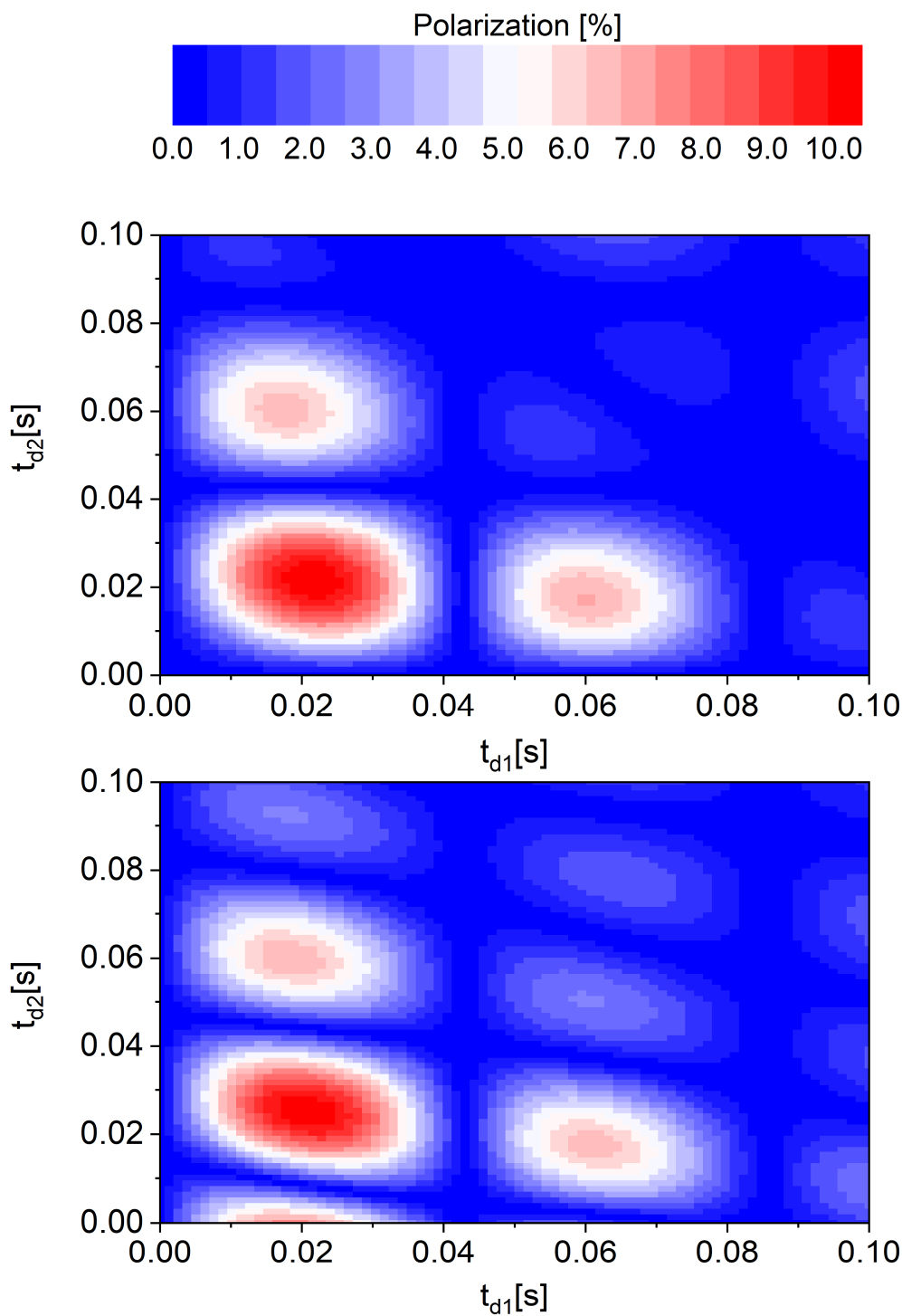

**Figure 6.** Simulations of the absolute bound-substrate polarization dependence of SABRE-INEPT on its free evolution delays $t_{d1}$ and $t_{d2}$. Parameters: $k_d = 10$, $\frac{[fS]}{[bS]} = 30$, $t_b = 10$ s, $T_1^{fS} = 60$ s, $T_1^{H_2} = T_1^{bS} = 1$ s. Top: singlet spin order of $H_2$. Bottom: anti-phase spin order of $H_2$.

## D. INEPT-type pulse sequences

Pulse sequences using coherence transfer have become an important tool in high-field SABRE. Here we demonstrate the capabilities of our model for the case of $^{15}N$ polarization by using INEPT-type sequences. Other applications, like the work of Tessari's group[24, 25, 31], can be treated as well by using the methods developed here. **Figure 6** shows the dependence of polarization generated by using the SABRE-INEPT sequence on the inter-pulse delays, $t_{d1}$ and $t_{d2}$. The maps have been calculated for the singlet spin order of H$_2$ and for the anti-phase spin order. One can see, that the dependences on both delay times contain oscillatory components, which decay in time due to relaxation and exchange. It is worth mentioning, that experimental results are in agreement with the case of anti-phase spin order (data not shown). For the simulations shown in **Figure 6** no delay between bubbling and application of the pulse sequence was introduced, which would surely decrease the achievable polarization levels.

Taking the SABRE-INEPT sequences as an example we explore the interplay of chemical exchange and relaxation of hydrogen at the SABRE complex (**Figure 7**). While **Figure 6** already shows that an increasing exchange rate interrupts the polarization transfer process of a growing number of complexes before it is completed, **Figure 7** shows that an exchange rate below optimal conditions will do the same, due to the fact that an increased lifetime of the complex means that the Ir-H$_2$ protons would relax before they can be replaced by fresh $p$H$_2$. Thus, the overall number of SABRE-active complexes is decreased. When relaxation of the Ir-H$_2$ protons is not considered, the efficiency of the INEPT-based approaches will increase towards the value one would obtain when optimizing only the spin dynamics.

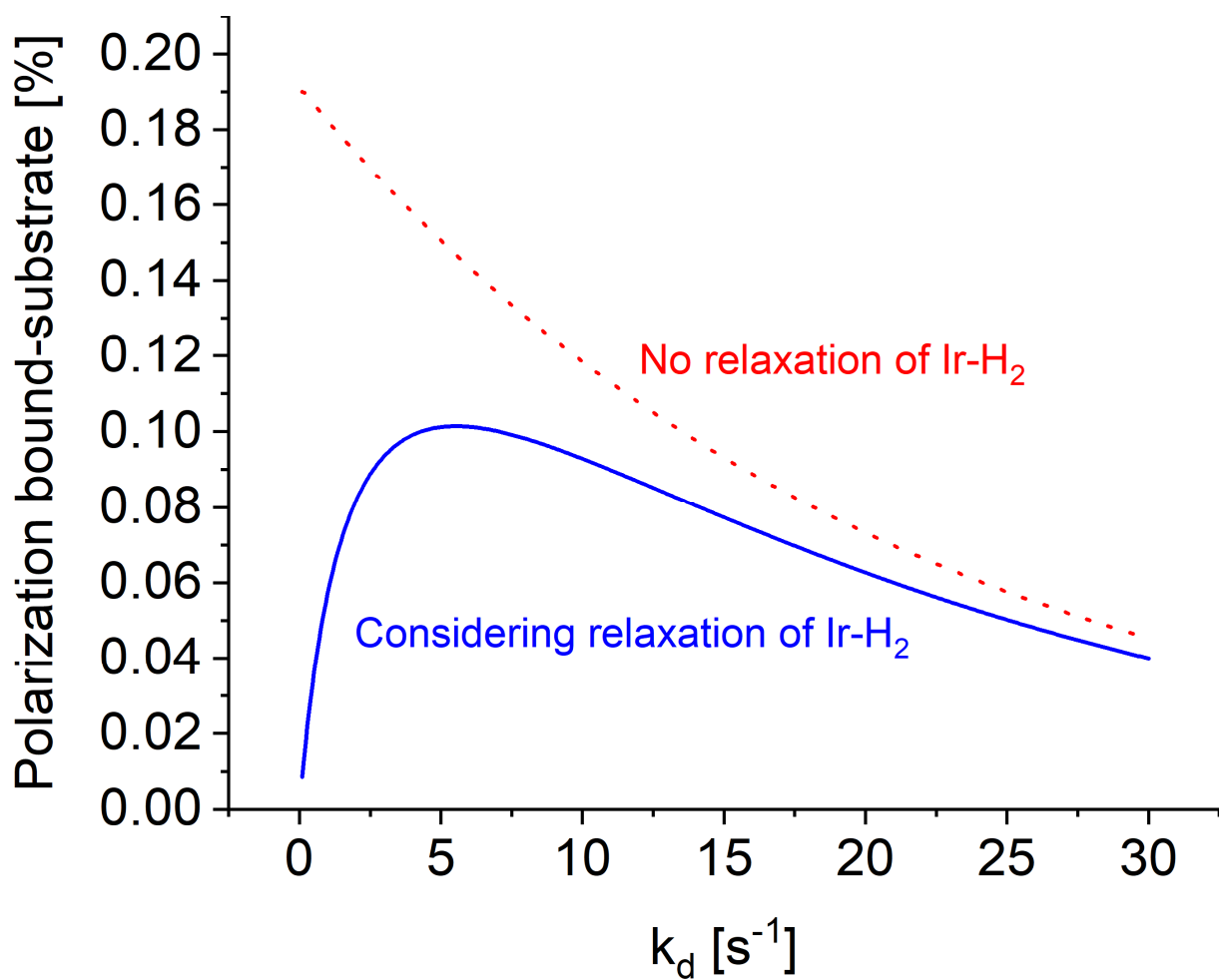

**Figure 7.** Polarization dependence of the bS on the dissociation rate of $k_d$ of the complex when the SLIC-SABRE sequence is applied. The dashed red line are simulations without relaxation of the Ir-H$_2$ protons. The blue solid curve shows the situation when relaxation is considered At the beginning both curves exhibit an increase in polarization as the $k_d$ is decreased, reflecting the fact that more complexes experience the full duration of the pulse sequence. When relaxation is considered the Ir-H$_2$ protons are relaxed more efficiently when their time in the complex is longer because of the relatively fast relaxation of the Ir-H$_2$ protons in the complex ($T^{IrH_2} = 1$ s).

The INEPT-type sequence we study here is also a useful example to investigate how polarization is spread to the free substrate pool. **Figure 8** shows simulated buildups of polarization of the fS and bS pools at various concentrations of fS. At time $t = 0$ only the bS polarization is present, owning to the fact that polarization is only generated in the SABRE complex. Subsequently, the polarized spins are distributed among the free and bound form and polarization levels equilibrate. One may notice, that although an equal ratio of free to bound substrate will yield the highest polarization in **Figure 8**, the overall signal generated is the same for all concentrations. Furthermore, the increased decay of free substrate with increasing fS concentrations is clearly visible.

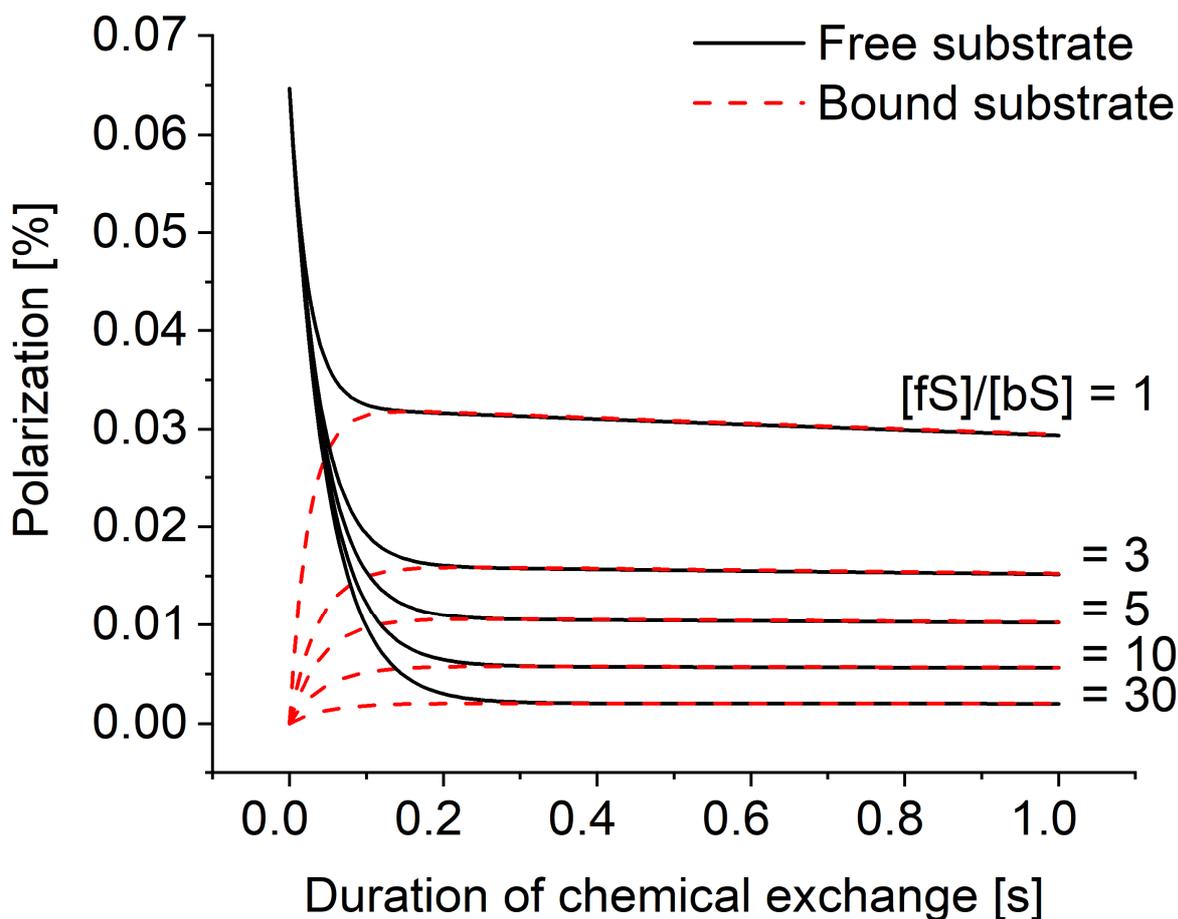

**Figure 8**. Simulations of the free-substrate polarization dependence of RF-SABRE on RF frequency $\omega_{rf}$ and amplitude. Parameters: $k_d = 20$, $\frac{[fS]}{[bS]} = 30$, $t_b = 60$ s, $T_1^{fS} = 60$ s, $T_1^{H_2} = 1$ s, $T_1^{bS} = 3$ s.

## V. Conclusions

We are able, for the first time, to make realistic predictions for the efficiency of high-field SABRE methods: previous considerations were limited to the treatment of the spin dynamics in the SABRE complex being unable to describe the interplay between the spin dynamics and chemical kinetics. We have analyzed existing methods and demonstrated the capabilities of our model. We have taken into account the effects, which are introduced by singlet-triplet mixing in dihydrogen and decay of the $H_2$ spin order. Furthermore, we are able to treat and analyze the effect of key NMR parameters and the SABRE kinetics on the resulting NMR signal enhancement. This is essential to get a realistic description of the behavior of RF-based transfer schemes and to optimize their performance in terms of polarization time and resulting signal enhancement. Our consideration allows us to estimate the performance of the existing polarization transfer schemes, which is summarized in **Table 2**. The simulations presented in this work show clearly, that such an estimation is hardly possible by simply providing an optimal polarization level, as it depends on a multitude of parameters. Thus, we provide a range of parameters, except for the case of LIGHT-SABRE, where near zero polarization is expected for the case of anti-phase spin order, while for the singlet spin order the polarization level should be close to that[13, 14] found in experiments using polarization transfer at ultralow fields.

**Table 2**. Performance of the transfer schemes: theoretical value in the limiting case of $H_2$ prepared in the pure singlet state; theoretical value for the anti-phase order of $H_2$; experimentally achieved values. Here we present results for fS polarization; substrate exchange is explicitly taken into account.

| Method | Theoretical singlet order of $H_2$ | Theoretical anti-phase order of $H_2$ | Experimentally achieved |
| --- | --- | --- | --- |
| RF-SABRE | 1-20 % | 0.1% - 1% | 0.5% |
| LIGHT-SABRE | 75% | < 1 % | ~0.1 % |
| INEPT | 1-20% | 1-20 % | ~1 % |

One can see that simulations, which do not take into account the actual spin order of $H_2$ (which is typically the anti-phase rather than the singlet order), strongly overestimate the performance of high-field SABRE techniques. It is virtually impossible to rationalize this discrepancy without a full theoretical treatment that takes all effects underlying the complex process of polarization formation. At the same time, simulations performed for the anti-phase order provide much more realistic estimates of the NMR enhancement. It is also interesting to note that some methods[20, 22] do not exploit the difference in the $|S\rangle$ and $|T_0\rangle$ populations thus being insensitive to $S$-$T_0$ mixing in $H_2$. Another possibility to tackle problems originating from $S$-$T_0$ mixing is partial recovery of the spin order by applying a pulse on the proton channel[18]. Another possibility for improving the performance of high-field SABRE is either keep singlet state alive somehow or adjust methods to new and more realistic conditions.

Even though we treat the degree of singlet-triplet mixing as an empirical external parameter, in principle, we could introduce differential equations, see e.g. eq. (7), to describe the influence of hydrogen exchange with intermediate complexes. We believe that in the future, when we have more information on these intermediates and their NMR properties, such models may be useful in predicting the behavior of the spin dynamics of this complex chemical process.

Hence, we anticipate that the proposed model should be useful to simulate more complex experiments and could be easily integrated into openly available NMR simulation software; We are also planning to compare different high-field SABRE schemes under identical experimental conditions in order to find out, which methods are preferable.

# Acknowledgements

This work has been supported by the Russian Science Foundation (grant No. 14-13-01053).